\documentclass[superscriptaddress,prd,twocolumn]{revtex4}

\usepackage{amstext,amsmath,amssymb,amsfonts,bbm}
\usepackage{amsthm}
\usepackage[dvips]{graphicx}
\usepackage{latexsym}
\usepackage{euscript}
\usepackage{epsfig}
\usepackage{psfrag}
\usepackage{ulem}

\theoremstyle{definition}
\newtheorem{theo}{Theorem}

\newtheorem{prop}{Proposition}

\newcommand{\C}{\mathbb{C}}

\DeclareMathOperator{\tr}{tr}

\def\be{\begin{equation}}
\def\ee{\end{equation}}
\def\bes{\begin{eqnarray}}
\def\ees{\end{eqnarray}}

\def\6{\langle}
\def\9{\rangle}

\def\hh{{\cal H}}
\def\cc{{\cal C}}
\def\ww{{\cal W}}

\def\6{\langle}
\def\9{\rangle}
\def\tr{{\rm tr}\,}
\def\half{\mbox{$1\over2$}}
\def\1{{{\mathbbm 1}}}

\begin{document}
\title{ Quantum causal histories in the light of quantum information}
\author{ Etera R. Livine}\email{etera.livine@ens-lyon.fr}
\affiliation{Laboratoire de Physique, ENS Lyon, CNRS UMR 5672, 46 All\'ee d'Italie, 69364 Lyon Cedex 07, France}
\author{ Daniel R. Terno}\email{dterno@perimeterinstitute.ca}
\affiliation{Perimeter Institute, 31 Caroline St, Waterloo, Ontario, Canada N2L 2Y5}

\begin{abstract}

We use techniques of quantum information theory to analyze the
quantum causal histories approach to quantum gravity. We show that
while it is consistent to introduce closed timelike curves (CTCs),
they cannot  generically  carry independent degrees of freedom.
Moreover, if the effective dynamics of the chronology-respecting
part of the system is linear, it should be completely decoupled
from the CTCs. In the absence of a CTC not all causal structures
admit the introduction of quantum mechanics. It is possible for
those and only for those causal structures that can be represented
as quantum computational networks. The dynamics of the subsystems
should not be unitary or even completely positive. However, we
show that other commonly maid assumptions ensure the complete
positivity of the reduced dynamics.
\end{abstract}
\maketitle



\section{Introduction}

The quest for a quantum theory of gravity produced a variety of
approaches that include string theory, loop quantum gravity, spin
foams, causal sets, and causal dynamical triangulations.  The
successful theory should provide a coherent structure that
accommodates both classical relativity and quantum mechanics, show
that the familiar physical phenomena on the flat spacetime
background emerge in some appropriate limit and finally make
predictions on the kind and magnitude of the departures from this
picture.

This final goal has not yet been achieved by any of the approaches,
but each of these attempts has brought many insights and led to a
better understanding of the problem's complexity. Quantum causal
histories (QCHs) approach \cite{fotini03} to the quantization of
gravity is a background-independent formalism that satisfies   many
of the conditions that are argued  for by the above models. The idea
is to use a causal set to describe the casual structure while a
quantum theory being introduced through the assignment of
finite-dimensional Hilbert spaces to the elementary events.
Originally motivated by quantum cosmology \cite{fotini00}, and
providing a description of the  causal spin foam models
\cite{causalSF}, QCHs make a direct contact with quantum computation
and quantum information theory \cite{nc} in general.

Quantum computation can be thought of as a universal theory for
discrete quantum mechanics. Quantum computers are discrete systems
that evolve by local interactions, and every such system can be
simulated efficiently on a quantum computer \cite{nc,lloyd}.  The
approaches to quantum gravity, and QCH in particular, depict it as
a discrete and  local quantum theory. Hence it should be
describable as a quantum computation \cite{lloyd}.

We apply the quantum-informational considerations to the several
questions in QCH. First, we consider closed timelike curves and
the modifications in quantum mechanics that they may cause. Next,
we show that the only causal histories that are compatible with
quantum mechanics are those that can be represented as quantum
computational networks. Finally we deal with the evolution of the
subnetworks and the role of completely positive maps.

The remainder of this section is devoted to the review of the
necessary concepts. We begin with a brief outline of QCHs in the
Hilbert space language, roughly following \cite{fotini03}. If a
spacetime is time-orientable and has no closed timelike curves,
then its causal structure can be completely described as a partial
order relation on its points. The relation $x\preceq y$ is defined
if there exists a future-directed non-spacelike curve from $x$ to
$y$. It is transitive, and the absence of CTCs means that
$x\preceq y$ and $y\preceq x$ are simultaneously true if and only
if $x = y$. Those two conditions make the relation ``$\preceq$"
into a partial order.

A discrete analogue of a smooth chronology-respecting spacetime is
a causal set $\cc$, which is a locally finite and  partially
ordered set. That is, for any two events $x, y\in\cc$, there exist
(at most) finitely many events $z\in\cc$ such that $x\preceq
z\preceq y$. If the events $x$ and $y$ are not related, i.e.,
neither $x\preceq y$ nor $y\preceq x$ holds, then   they are
spacelike separated, this fact being denoted as $x\sim y$. At the
discrete level there is no distinction between causal and
chronological
\cite{gr} entities.
An acausal set is a subset $\xi\in\cc$ such that all events in it are spacelike separated from one
another. Then maximal acausal sets are the discrete analogues of spacelike hypersurfaces.

\begin{figure}[htbp]
\epsfxsize=0.52\textwidth
\centerline{\epsffile{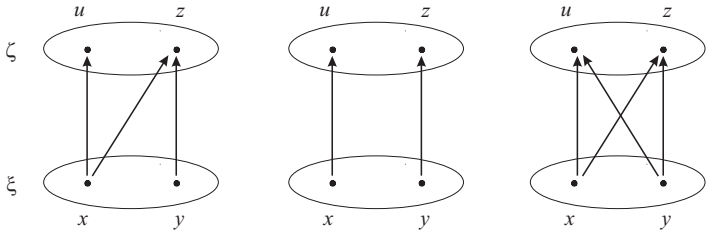}}
\caption{\small Three possible causal histories}\label{3type}
\end{figure}

A causal set can be represented by the directed graph of
elementary relations, as on Fig.~\ref{3type}. Its vertices are the
points of $\cc$, while the edges $x\rightarrow y$ represent the
elementary causal relations, namely $x\preceq y$ without any
intermediate $z$ such that $x\preceq z\preceq y$.

A future-directed path is  a sequence of events such that there
exists an edge from each event to the next. It is an analogue of a
future-directed non-spacelike curve. A future-directed path is
future (past) inextendible if there exists no event in $\cc$ which
is in the future (past) of the entire path.  Then one can define
complete future and complete past of an event.  An acausal set
$\xi$ is a complete future of an event $x$ if $\xi$ intersects any
future-inextendible future-directed path that starts at $x$, and a
complete past is defined similarly. If an acausal set $\zeta$ is a
complete future of an acausal set $\xi$ and at the same time the
set $\xi$ is a complete past of $\zeta$, then the sets form a
complete pair, $\xi\preceq \zeta$.

 A local quantum structure on
a causal set is introduced by attaching  a finite-dimensional
Hilbert space $\hh(x)$ to every event $x\in\cc$. For two spacelike
separated events $x$ and $y$ the composite state space is
$\hh(x,y)=\hh(x)\otimes\hh(y)$, with an obvious generalization to
larger sets. In ordinary quantum mechanics (of closed systems)
time evolution is a unitary map of Hilbert spaces. In a QCH
approach one introduces a unitary evolution between complete pairs
of acausal sets $\xi$ and $\zeta$, where, e.g., $\zeta$ is the
complete future of $\xi$, $\xi\preceq\zeta$. One can think of a
complete pair as successive Cauchy surfaces of an isolated
component of spacetime, or of all spacetime, with a unitary map
$U$ relating $\hh(\xi)$ and $\hh(\zeta)$.

Hence a QCH consists of a causal set $\cc$, a finite-dimensional
Hilbert space $\hh(x)$ at every $x\in\cc$ and a unitary map
\be
U(\xi,\zeta):\hh(\xi)\rightarrow \hh(\zeta) \label{defu}
\ee
 for any complete
pair $\xi\preceq\zeta$. The maps have a natural composition
property
\be
U(\varsigma,\zeta)U(\xi,\varsigma)=U(\xi,\zeta), \qquad {\rm for~}
\xi\preceq \varsigma\preceq\zeta.
\ee
Different possible causal relations between the complete sets are
shown on Fig.~\ref{3type}.

\begin{figure}[htbp]
\epsfxsize=0.33\textwidth
\centerline{\epsffile{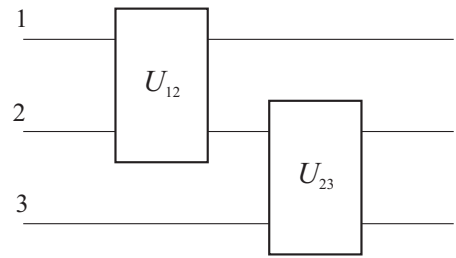}}
\caption{\small A quantum circuit with two gates}\label{circ}
\end{figure}

Quantum circuits \cite{nc}, as the one depicted on
Fig.~\ref{circ}, represent sequences of unitary operations (that
are called gates) that are performed on one or several quantum
wires, that represent distinct discrete quantum systems. Usually
those  systems are qubits --- two-dimensional quantum systems. The
wires carry information around the circuit, and the conventional
direction from left to right may correspond to the passage of
time, or to the information carrier moving from one location to
another. Quantum circuits do not contain loops, so there is no
feedback from one part of the circuit to another. Quantum states
cannot be cloned, so the wires do not split up. There are special
symbols for particular set of gates, but through this paper we use
only a generic $n$-partite box symbol.

Evolution of an open quantum system is non-unitary. When the initial correlations with the
environment can be ignored, the resulting dynamics is completely positive (CP) \cite{nc,rmp,ctz}.
This is a crucial property: a linear map $T(\rho)$ is called positive if it transforms any positive
matrix $\rho$ (namely, one without negative eigenvalues) into another positive matrix.  It is
called completely positive if $(T\otimes\1)$ acting on a bipartite $\rho$ produces a another
bipartite state, i.e., a positive trace-one operator. For instance, complex conjugation of $\rho$
(whose meaning is time reversal) is a positive map, since it preserves the eigenvalues of the
Hermitian matrix. However, it is not completely positive and as such is used to identify entangled
states \cite{nc,rmp}. CP maps are the integral part of the toolbox of quantum information. Later we
discuss their role in QCHs.

The rest of this paper is organized as follows. In the next
Section we discuss CTCs. Section~III deals with the possible types
of the causal relations that permit to introduce quantum
mechanics. Evolution of subsets of the complete pairs is subject
of Section~IV. In Section~V we discuss the relation between local
and global information about quantum states.

\section{Structure: absence of causal loops}
 Solutions of
Einstein equations with CTCs have been known for a long time, and they re-emerged to the public eye
after introduction of traversable wormholes \cite{worm,fn}. A discrete setting
\cite{deutsch} makes it easier to analyze potential paradoxes that
result from the  presence of CTCs.

The basic assumption in construction of a QCH is that the set $\cc$ is partially ordered, i.e. it
contains no CTCs. Which kind of quantum mechanics, if any, results from their introduction depends
on the definition of the discrete analogues of the equal time surfaces. We show that allowing
interactions between the systems in the chronology-respecting and chronology-violating regions
restricts the allowed CTCs, regardless of  other assumptions. The rest of the properties, both from
the point of view of a local observer and a global ``superobserver", depend on the additional
hypotheses that are described below.

\begin{figure}[htbp]
\epsfxsize=0.4\textwidth
\centerline{\epsffile{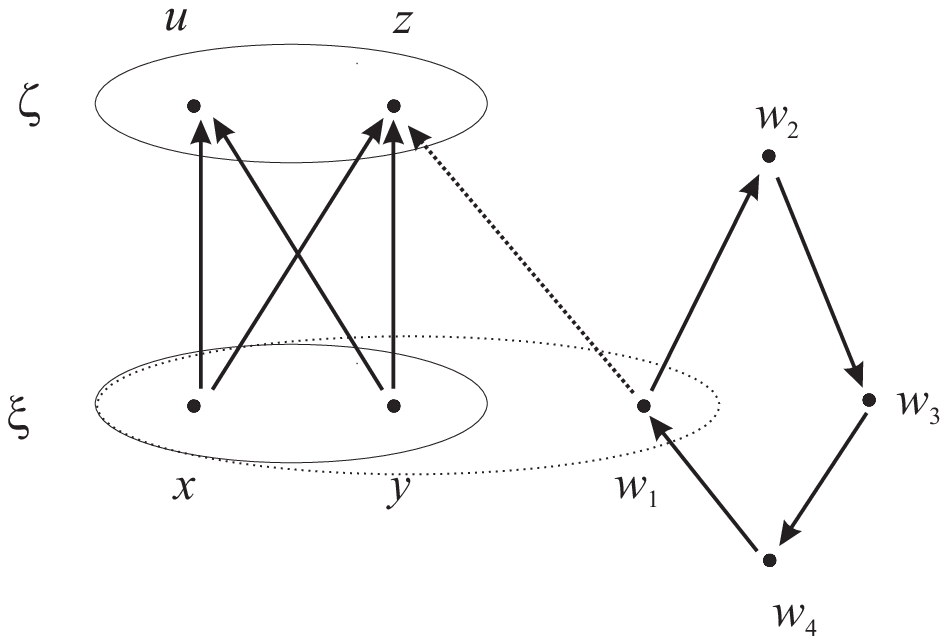}}
\caption{\small The complete past of the set $\zeta=\{u,z\}$} is $\xi=\{x,y\}$ or
$\bar{\xi}=\{x,y,w_1\}$, if there is a causal relation between
$\ww$ and $\zeta$.\label{loopjoin}
\end{figure}

 We use Fig.~\ref{loopjoin} to
illustrate  possible definitions. The diagram contains a CTC $\ww$
that is made of the four-point set $W$, $w_1,\ldots w_4\in W$ and
the edges between them. This is a chronology-violating set, since
both statements $w_i
\preceq w_j$ and $w_i
\succeq w_j$ are true for all points of $W$.

There are two possible ways in which $\ww$ can be embedded into a
larger diagram. If there are no causal relations between the
normal (chronology-respecting) region and the points of the CTC,
then it is disconnected from the rest of the set and can be simply
ignored. On the other hand,  points of the CTC may be allowed to
influence the chronology-respecting region, and in may be
influenced by it, similarly to the continuous case
\cite{fn}.

\begin{figure*}[htbp]
\epsfxsize=0.4\textwidth
\centerline{\epsffile{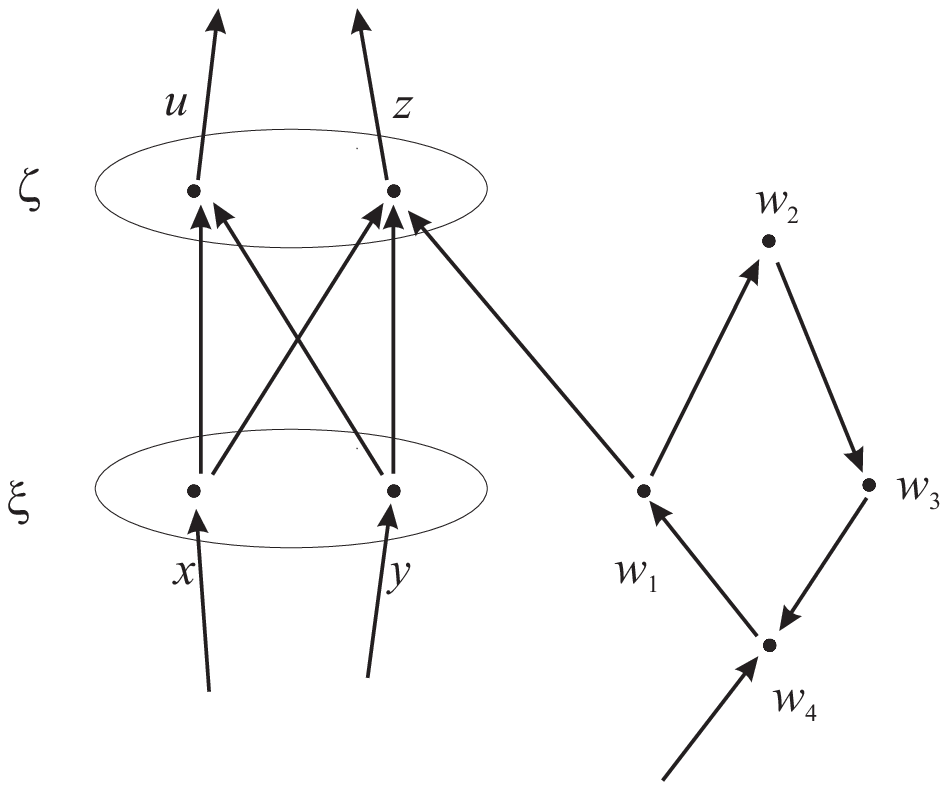}\hspace{0.8cm}\epsfxsize=0.4\textwidth\epsffile{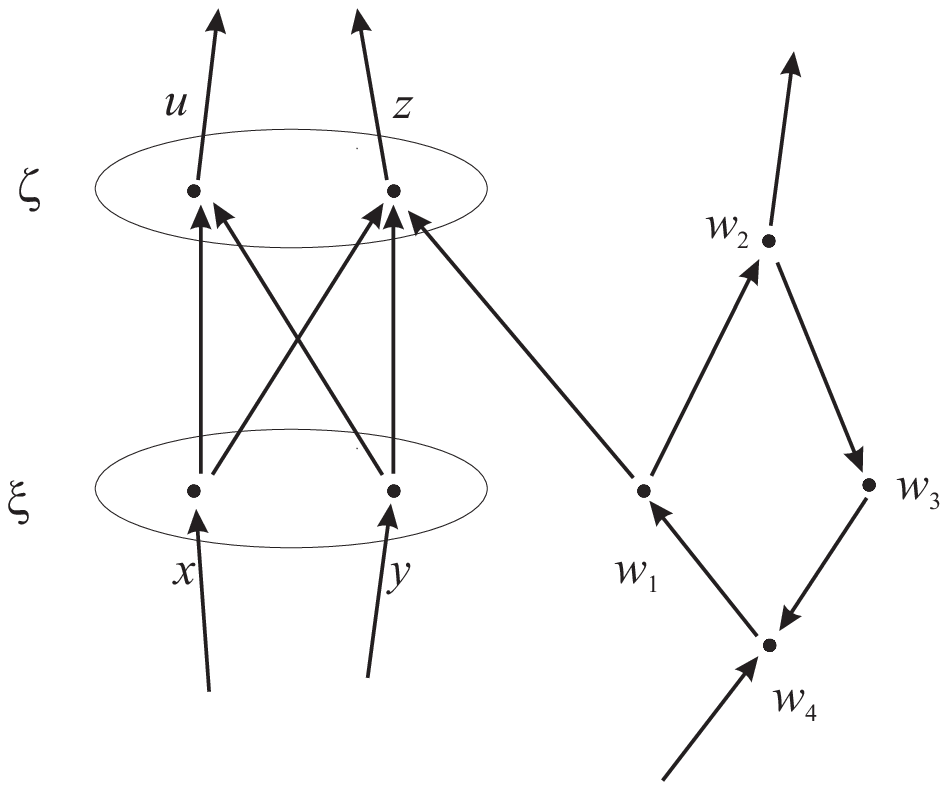}}
\caption{\small{Two ways of embedding the diagram of Fig.~\ref{loopjoin} into a bigger set}.}\label{loopbig}
\end{figure*}

The minimal extension of the rules is to allow  points of a CTCs
into an acausal set  if they are causally disconnected from the
rest  of the set. Fig.~\ref{loopjoin} represents a part of one
possible causal structure. Since any single point of $\ww$ is
spacelike separated from the points of $\xi$, a set
$\bar{\xi}=\xi\cup\{w_i\}$ is also an acausal set. Similarly,
other acausal sets are $\bar{\xi}'=\xi\cup\{w_2\}$,
$\varsigma=\{u,w_2\}$, etc. The loop $\ww$ is not
future-inextendible, since $z\succeq\ww$. As a result, the
complete future of $w_1$ (or any other point of $\ww$) is an
acausal set $\zeta$. To avoid the question  of how to interpret
the definition of the past-inextendible path, assume that the
Fig.~\ref{loopjoin} is a part of a larger causal structure that is
given on Fig.~\ref{loopbig} (a).

\begin{figure}[htbp]
\epsfxsize=0.35\textwidth
\centerline{\epsffile{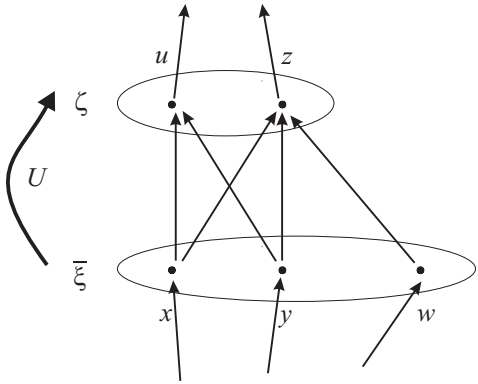}}
\caption{\small Effective causal structure that disregards the CTC altogether.}\label{loopeffective}
\end{figure}

It is easy to see that the set $\xi$ is the complete past of
$\zeta$, hence the two sets form a complete pair. When we
introduce quantum mechanics, the  rest of the points of $\ww$ are
irrelevant, and the effective diagram is represented on
Fig.~\ref{loopeffective}. On the other hand, the point $w_1$ (and
together with it the set $\bar\xi$) on Fig.~\ref{loopbig} (b) has
no complete future at all, since there is no acausal set that
intersects both future-directed future-inextendible paths that
start at $w_1$. In this case introduction of quantum mechanics as
described in Section I is impossible.

Hence if the complete pairs of the (generalized) acausal sets are
necessary to introduce quantum mechanics, only CTCs that are
compatible with the existence of such pairs are allowed. In this
case
 a single representative of the loop is picked and
treated as  part of a standard partially ordered causal structure,
and the existence of CTCs has no consequence.

It looks more natural to relax a demand of the acausality
 in the definitions of the  past and future sets \cite{deutsch,bac}.
 One
allows a single point from each CTC  into such an ``acausal" set,
so the points may be  causally related  only through the loop. For
 example, the set
$\bar{\zeta}=\zeta\cup\{w_2\}$ may be taken to be the complete
future of $\bar{\xi}$, which is then its complete past.

 To avoid the inconsistency of
quantum theory in this model  one must impose  a
 self-consistency requirement
\cite{deutsch}. Since  the points of $\ww$ belong to the complete
past of $\zeta$, the (reduced) state on all events $w_i$ should be
the same.
 That means, if
$\rho_w$ is the state in the CTC region at the ``temporal origin''
at $w_1$ it should be the same ``after'' the evolution $U$.

Since the preparation procedure is possible at any point of the
chronology-respecting region, the state on $\hh=\hh_A\otimes\hh_B$
is taken to be a direct product state
\cite{deutsch}. Here $\hh_A$ stands for the Hilbert space of the chronology-respecting set, and $\hh_B$ for the CTCs.
Then the evolution (e. g., $U:
\hh(\bar\xi)\rightarrow\hh(\bar\zeta)$ of the above example) is supplemented by the
consistency condition
\be
\rho_B=\tr_A[U(\rho_A\otimes\rho_B)U^\dag]. \label{causcon}
\ee
 It
was shown
 that there is always at least one solution for this
self-consistency equation and in some cases  $\rho_B$ belongs to a
continuous family of the solutions \cite{deutsch,bac}. However, we
now show that for generic $U$ and $\rho_A$ the solution is unique.

Any state $\rho$ on $d_A\times d_B$ dimensional space
$\hh=\hh_A\otimes\hh_B$ can be decomposed as
\bes
\rho=\frac{1}{d_Ad_B}(\sigma^0_A\otimes\sigma^0_B+\sum_i\alpha_i\sigma_A^i
\otimes\sigma^0_B+\nonumber\\\sum_j\beta_j\sigma^0_A\otimes\sigma_B^j
+\sum_{i,j}\gamma_{ij}\sigma_A^i\otimes\sigma_B^j),
\label{Fano}
\ees
Here $\sigma^i_X$ represent generators of SU($d_X$),
$\sigma^0=\1$ and the real vectors $\vec \alpha$ and $\vec \beta$ of
the size $d_A^2-1$ and $d_B^2-1$, respectively, are  the generalized
Bloch vectors of the reduced density operators. If the state is a
direct product $\rho_A\otimes\rho_B$, then
$\gamma_{ij}=\alpha_i\beta_j.$ Denote the action of $U$ as
\bes
U\sigma^\alpha_A\otimes\sigma_B^\beta U^\dag =\sum_{\mu,\nu}
s^{\mu\nu}_{\alpha\beta}\sigma^\mu_A\otimes\sigma_B^\nu,\nonumber\\
\alpha,\mu=0,\ldots d_A^2-1, \qquad \beta,\nu=0,\ldots d_B^2-1.
\ees
The consistency condition is a linear system
\be
\sum_i\beta_i(\delta_{im}-s_{0i}^{0m}-\sum_j\alpha_js_{ji}^{0m})=\sum_i\alpha_i
s_{i0}^{0m}.\label{system}
\ee
Its solution is not unique  if and only if
\be
\Delta({\vec
\alpha})=\mathrm{Det}\|\delta_{im}-s_{0i}^{0m}-\sum_j\alpha_js_{ji}^{0m}\|=0
\ee
To have a $\Delta(\vec{\alpha})$ for a generic $\rho_A$ (apart
from a finite set of states), two equations should be satisfied
simultaneously:
\be
\mathrm{Det}\|\delta_{im}-s_{0i}^{0m}\|=0,
\ee
and
\be
\sum_j\alpha_js_{ji}^{0m}=0.
\ee
Since the last equation  holds for all but a finite set of
$\alpha_j$ it means that
\be
s_{ji}^{0m}=0, \qquad \forall j=1,\ldots,d_A^2-1
\ee
Hence the solution of a self-consistency equation is not unique
only for a subset of lower dimensionality of the set of all
unitaries on $\hh=\hh_A\otimes\hh_B$.

This points to a  crucial difference  with the usual quantum
theory. The  Hilbert space $\hh_B$ on a CTC generically carries no
independent degrees of freedom: the states $\rho_B$ are uniquely
determined from $U$ and $\rho_A$, apart from the set of a measure
zero.

From Eq.~(\ref{system}) it follows that $\rho_B$ is a rational
function of $\rho_A$ and so a generic $U$
\be
\rho'_A=\tr_B[U(\rho_A\otimes\rho_B)U^\dag],
\ee
leads to a non-linear evolution. Actually, from the point of view
of a local observer to whom only the chronology-respecting set is
accessible,  any evolution other than $U_A\otimes\1_B$ produces a
non-linear local dynamics.   Writing explicitly
\begin{widetext}
\be
\rho'_A=\frac{1}{d_A}\left(\sigma^0_A+\left(\sum_i\alpha_is_{i0}^{m0}+\sum_j\beta_j(\vec{\alpha})s_{0j}^{m0}+
\sum_{ij}\alpha_i\beta_j(\vec{\alpha})s_{ij}^{m0}\right)\sigma^m_A\right),
\ee
\end{widetext}
one sees that since for a generic $\rho_A$ the consistent $\vec{\beta}$ is a non-linear function of
$\vec{\alpha}$, the linear evolution on $\hh_A$ is possible only if
\be
s_{0j}^{m0}=0,\qquad s_{ij}^{m0}=0, \qquad \forall j=1,\ldots
d^2_B-1.
\ee As a
result, the state of $\hh_A$ is independent of the ``environment",
hence its unitary evolution is of the form $U_A\otimes\1_B$.

Ignoring a ``conspiracy" question about a mechanism that prepares
the state $\rho_B$ in accordance with the state $\rho_A$ in the
causally disconnected region, the non-linearity of the resulting
evolution is an observable consequence of a non-trivial
interaction with CTC states.

The consistency condition also restricts the number of causal
links that a CTC can have with the chronology-respecting part of
the diagram. As it was argued above, in the generic case the state
of the loop is uniquely determined by $\rho_A$ and $U$. That means
if there is a second causal link with the loop, the consistency
condition now becomes
\be
\rho_B=\tr_A[V\rho_{AB}V^\dag]=\tr_A[VU\rho_A\otimes\rho_B U^\dag V^\dag],
\ee
which holds for for the set of measure zero of the bipartite
unitary transformations or may have no solution at all. Even when
the state of the loop is not determined uniquely, every imposition
of the consistency condition reduces the dimensionality of the set
of consistent states $\rho_B$ at least by 1. Hence, if we are
ready to accept the lower dimensional set of the possible
evolutions, a CTC cannot have more that $d_B^2-1$ links with the
chronology-respecting part of the graph, while if one insists on
generic $U$, there could be only a single link.

We  conclude the following. Within the strict interpretation of the
acausal surfaces, CTCs either prevent the introduction of quantum
theory altogether, or lead to no (observable) changes by
contributing an extra subspace to the ordinary chronology-respecting
system. Under the relaxed rules \`a la Deutsch \cite{deutsch}, the
demand of linearity of the dynamics leads to a decoupling of a
chronology-respecting region and  CTCs. If a non-linearity is
accepted, then the CTC region does not carry independent degrees of
freedom and only a certain amount of causal relations between the
regions may be allowed.

\section{Quantum Histories and the Observer-Dependence of Causal Links}

From now we will work on a causal set $\cc$, which has no timelike
loops. Nevertheless, it turns out that not all causal histories
are consistent with quantum mechanics. Note that the precise
meaning of arrows in the relation $x\preceq y$ is as follows. For
$x\in\xi$ and $y\in\zeta$ the existence of $U(\xi,\zeta)$ implies
that if there is no future-directed path between $x$ and $y$,
i.e., $x\sim y$, then the reduced final state on $\hh(y),$
$\rho^\zeta_y=\tr_{\zeta\!\setminus
y}\rho^\zeta=\tr_{\zeta\!\setminus y}U\rho^\xi U^\dag$ is
independent of the initial reduced state $\rho^\xi_x$.

We deal first with the situation when it is possible to identify
the initial and final Hilbert spaces pointwise, e.g., on
Fig.~\ref{3type} we assume that
$d_x\equiv\dim\hh(x)=d_u\equiv\dim\hh(u)$, etc. A general case is
considered at the end of this section.

 Consider three possible causal relations
that are presented on Fig.~\ref{3type}. Despite its intuitive
appeal the causal history (a) is incompatible with the defining
Eq.~(\ref{defu}). It can be observed on a simple example of two
qubits. Label the basis of each of the spaces by $|0\9,|1\9$. The
controlled-NOT (CNOT) unitary operation on two qubits
\cite{nc} apparently fits the described scheme: the value of the
qubit $x$ remains the same, while the qubit $y$ may be flipped.
The action of CNOT on this basis is given in the first two columns
of the table below.
 However, in the basis $
|\pm\9=(|0\9\pm|1\9)/\sqrt{2}$, the same unitary evolution should
be depicted with the arrow going from $y$ to $x$.

\begin{center}\begin{tabular}{|c|c||c|c|}\hline
$|\psi\9$ & $U|\psi\9$ &$|\psi\9$ & $U|\psi\9$ \\\hline
00 & 00 & ++ & ++ \\
01 & 01 & $-+$ & $- +$ \\
10 & 11 & $+ -$ & $- -$ \\
11 & 10 & $- -$ & $+ -$ \\
\hline
\end{tabular}
\end{center}

In general if an operation on $\hh(y)$ is controlled by the state
on $\hh(x)$ which remains unchanged,
\be
U(|\psi\9\otimes|\phi\9)=|\psi\9\otimes V_\psi|\phi\9,
\label{form}
\ee
for all states $|\phi\9$, then it is actually independent of
$|\psi\9$, $V_\psi\equiv V$. Indeed, consider two possible initial
states $|\psi\9|\phi\9$ and $|\psi'\9|\phi\9$. The overlap is
preserved under $U$, hence
\be
\6\phi|V_\psi^\dag V_{\psi'}|\phi\9=1,
\ee
for  an arbitrary state $|\phi\9$. Hence $V_\psi= V_{\psi'}$.

As a result, only the second and the third histories of Fig.~\ref{3type} are consistent with
existence of a unitary evolution on causal sets. The alternative is to  introduce an external
(classical) observer who is restricted to do measurements in a (given) particular basis, say (0,1)
as in the above example. The CNOT example shows that the observer will prescribe different causal
relations depending on its choice of measurement basis. Then, by restricting the allowed unitary
evolutions between the complete sets, the asymmetric causal structure is made compatible with
quantum mechanics.

Adhering to the latter option is not only too restrictive, but not always possible. Consider the
projectors $P_0$ and $P_1$, $P_0+P_1=\1$, on the singlet (spin-0) and triplet (spin-1) states of
 $\C^2\otimes\C^2$.
Then for generic values of the parameters $\alpha$, $\beta$ there
is no basis in which the unitary
\be
U=e^{i\alpha}P_0+ e^{i\beta}P_1,
\ee
can be represented as in Eq.~(\ref{form}). This particular set of unitary operators is relevant to
computational universe models \cite{lloyd} and to loop quantum gravity \cite{OHrep}.

\begin{figure}[htbp]
\epsfxsize=0.5\textwidth
\centerline{\epsffile{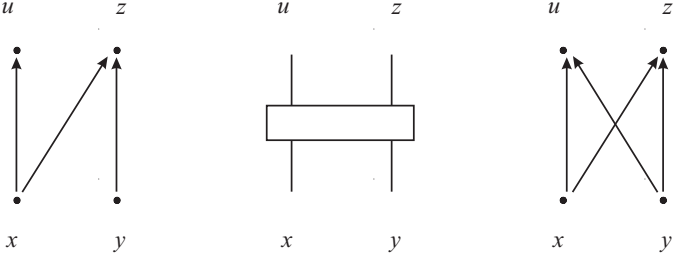}}
\caption{\small Two consecutive duality transformation}\label{dual}
\end{figure}

From the above discussion it follows that we have to work with the symmetric diagrams that take
into account mutual influence of quantum systems. This is automatically taken into account in the
dual picture, where edges represent quantum systems and vertices the interactions. In a graphic way
it is obtained by turning points of the causal set diagram into lines, and the arrows between the
points into boxes that cover the lines.   Applying the same transformation once more, we obtained a
symmetrized version of the initial diagram. For example, if one starts from the diagram (a) on
Fig.~\ref{3type} two duality transformations bring it to the diagram (c) on Fig.~\ref{dual}. It is
the standard quantum network picture of quantum information theory
\cite{nc}. This remains true also for more complicated diagrams,
the complete surfaces of which are considered as bipartite
systems.

\begin{figure*}[htbp]
\epsfxsize=0.7\textwidth
\centerline{\epsffile{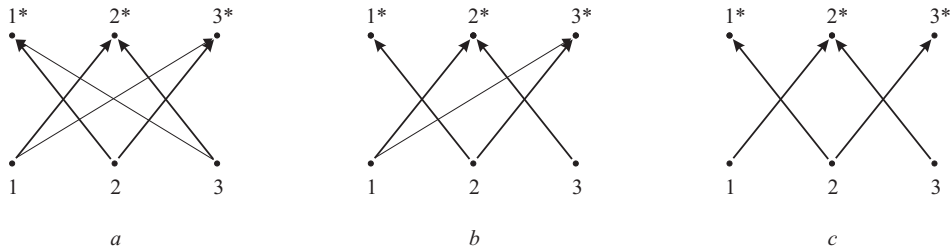}}
\caption{\small Possible causal links at the one step of the evolution of a tri-partite system. The trivial links such as $1\rightarrow 1^*$
are not shown. }\label{tri}
\end{figure*}

However, having three or more events in the complete surfaces allows for more complicated
structures, as shown on Fig.~\ref{tri}. The diagram (7b) and its mirror image with $1\sim 3^*,
3\preceq 1^*$ are consistent with the symmetry requirement. Any bipartite splitting (1 and 2 vs. 3,
etc.) results in a symmetric coarse-grained diagram.

\prop The diagram (7b) admits a unitary evolution $U$ if and only if
\be
U=U_{23}U_{12}, \qquad [U_{23},U_{12}]\neq 0,\label{twodec}
\ee
where $U_{23}$ and $U_{12}$ are the bipartite unitary operators
that act on $\hh_2\otimes\hh_3$ and $\hh_1\otimes\hh_2$,
respectively.

\medskip

First we prove the sufficient condition, namely that $1\preceq 3'$
and $3\sim 1'$. While from the circuit diagram of Fig.~\ref{circ}
that gives a graphic decomposition of  $U$ the latter statement is
obvious, it is worthwhile to consider a formal proof. Since linear
operators form a Hilbert space with a Hilbert-Schmidt inner
product, a bipartite unitary operator can be written as
\be
U_{AB}=\sum_\mu \alpha_\mu A_\mu\otimes B_\mu,
\ee
where the operators $\{A_\mu\}$ and $\{B_\mu\}$ form the
orthogonal bases,
\be
\tr A_\mu^\dag A_\nu=d_A\delta_{\mu\nu},\qquad \tr B_\mu^\dag
B_\nu=d_B\delta_{\mu\nu},
\ee
and the Schmidt coefficients satisfy
\be
\alpha_\mu>0,\qquad \sum_\mu|\alpha_\mu|^2=1,
\ee
and there is no more than $\min(d_A,d_B)$ terms. Hence
\be
U_{12}=\sum_\mu \alpha_\mu A_\mu\otimes B_\mu\otimes\1,\qquad
U_{23}=\1\otimes\sum_\nu C_\nu\otimes D_\nu.
\ee
It is enough to consider initial  pure product state \cite{rmp},
so
\be
U(|\phi\9_1\otimes|\varphi\9_2\otimes|\psi\9_3)=\sum_{\mu\nu}\alpha_\mu\beta_\nu
A_\mu|\phi\9_1\otimes C_\nu B_\mu|\varphi\9_2\otimes
D_\nu|\psi\9_3
\ee
The reduced density matrix
\bes
\rho'_1=\sum_{\mu\nu\mu'\nu'}\alpha_\mu\alpha_{\mu'}\beta_\nu\beta_{\nu'}\times\nonumber\\
\tr(B_\mu|\varphi\9\6\varphi|B_{\mu'}^\dag C_{\nu'}^\dag C_\nu)
\6\psi| D_{\nu'}^\dag D_\nu|\psi\9
A_\mu|\phi\9\6\phi|A_{\mu'}^\dag.
\ees
Expanding the trace and using $\6k|\6 l|U_{23}^\dag
U_{23}|k'\9|l'\9=\delta_{kk'}\delta_{ll'}$, we get
\be
\rho'_1=\sum_{\mu\mu'}\alpha_\mu\alpha_{\mu'}\6\varphi|B_{\mu'}^\dag
B_\mu|\varphi\9A_\mu|\phi\9\6\phi|A_{\mu'}^\dag,
\ee
as expected. On the other hand, since $[C_\nu,B_\mu]\neq 0$ at
least for some  $\mu$ and $\nu$,
\bes
\rho_3'=\sum_{\nu\nu'}\beta_\nu\beta_{\nu'}\6\varphi|C_{\nu'}^\dag
C_\nu|\varphi\9 D_\nu|\psi\9\6\psi|D_{\nu'}^\dag+\nonumber \\
\sum_{\mu\nu\mu'\nu'}\6\phi|A_{\mu'}^\dag
A_\mu|\phi\9\6\varphi|O(\mu,\mu',\nu,\nu')|\varphi\9D_\nu|\psi\9\6\psi|D_{\nu'}^\dag,
\label{compart}
\ees
where the operator $O$ is built from the Schmidt basis operators
and their commutators. Since this operator is non-zero, its
expectation is non-zero at least for some $|\varphi\9_2$, and the
state $\rho_3'$ indeed depends on $|\phi\9_1$.

To establish the necessary condition we show that any other form of $U$  corresponds to a different
diagram. The commuting $U_{12}$ and $U_{23}$ correspond to (7c), as shown below. Any tripartite
unitary can be decomposed \cite{nc} as certain products of the bipartite unitary operators. If it
is not of the form Eq.~(\ref{twodec}), then
\be
U=VU_{12}'U_{23}U_{12},
\ee
where $V$ is some unitary (which may be equal to the identity) and
$U_{12}'$ acts on $\hh_1\otimes\hh_2$. The relation $1\preceq 3^*$
follows from the the sufficient condition applied to
$U_{23}U_{12}$, and the relation $3\preceq 1^*$ is established by
the factor $U_{12}'U_{23}$. \hfill $\Box$

\prop The diagram (7c)   admits a unitary evolution $U$ if and only if
\be
U=U_{23}U_{12}, \qquad [U_{23},U_{12}]=0,\label{twocom}
\ee
The sufficient part follows from Eq.~(\ref{compart}), since in
this case $O=0$. Proposition~1 guaranties that any other form of
$U$ introduces  additional causal links. \hfill $\Box$

It should be noted that the diagrams with less-than-maximal number
of links correspond to the zero-volume subset of the set of all
unitary operators. For example,  a generic three-qubit unitary is
characterized by 64 parameters, while two two-qubit unitaries have
at most 32.

\begin{figure}[htbp]
\epsfxsize=0.2\textwidth
\centerline{\epsffile{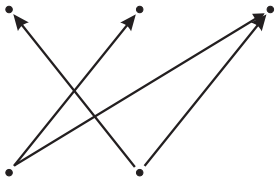}}
\caption{\small
In a generic causal history it is impossible to associate fixed Hilbert spaces to several
individual events (points). This leads to the natural question: how to define/characterize
individual events? }\label{223}
\end{figure}

 Fig.~\ref{223} gives an example of a graph where it is
 impossible to identify the spaces of the individual events.
 However, since all the spaces are finite-dimensional, their
 dimensions can be uniquely represented as a monomial
 $p_1^{n_1}\ldots p_k^{n_k}$, where the numbers $p_i$ are prime and
 the integers $n_i\geq 1$. Since the dimensionality of the
 Hilbert spaces on the complete surfaces are equal, it is possible
 to decompose them into fictitious subspaces of the dimensions that
 are prime powers. Those subspaces can be identified and the above
 consistency consistency requirements can be applied to them.

Our discussion established the following
\theo A QCH admits a unitary evolution between its acausal
surfaces if and only if it can be represented as a quantum
computational network.

\medskip

This quantum computational network, or more simply quantum circuit, describes the dual causal set:
events (points or boxes) are the unitary evolutions and objects/states are represented as arrows
(or lines) between the events. As we already saw in the example of the CNOT gate, this quantum
circuit picture allows a better representation of the causal relations. The bi-partite system case
was shown on Fig.~\ref{dual}. As for a tri-partite system as on Fig.~\ref{tri}, we have once again
a priori three lines coming in a big box representing the unitary evolution and then the same three
lines coming out. This picture can nevertheless be refined for the more peculiar causal structures
(7b) and (7c). Indeed we will now have a couple of 2-lines boxes, one for the evolution of the
system $1\otimes 2$ and one for the system $2\otimes 3$, as shown on Fig.~\ref{circ}. For the
diagram (7c), the order of these two boxes does not matter since the two unitaries commute with
each other. On the other hand, the history (7b) gives a priori a precise time ordering of the two
boxes, (12) coming before (23), and no change of measurement basis can lead to a causal arrow
between 3 and $1^*$.

\section{Completely positive maps}
 Unitary maps on the complete
pairs tell  very little about the causal relations between the two
acausal sets. To overcome this difficulty  completely positive
(CP) trace-preserving dynamics for states of the subsystems that
are associated with parts of complete surfaces of QCH was
postulated
\cite{fotini03, fotini05}. A trace-preserving map of density
matrices is dual to a unital (i.e., taking $\1$ to $\1$) map on
observables.

 A priori there is no reason to assume complete
positivity of the reduced dynamics. Indeed, it is known that prior
correlations (not necessarily entanglement) between the subsystems
may lead to a non-completely positive evolution. Hence enforcing
this requirement excludes many types of dynamics that establish
correlations between the subsystems. The requirement that the
reduced dynamics is unital, i.e., the maximally mixed state on
$\hh(x)$ is a fixed point of the map, is more natural
\cite{fotini03}. Indeed, an average over all CP
maps is the map $\Lambda$, which transforms the whole state space
into the total mixture,i.e., $\Lambda(\rho)\propto\1$. On the
other hand, in the absence of measurement data an arbitrary test
state at the output
 can be estimated as a total mixture. Consequently, the resulting
 reconstructed map is trivially unital \cite{jeanes, buzek}.

The reduced dynamics can be discussed as follows. Consider a
complete pair of acausal sets that are related by a unitary
evolution $U$ and let a part of the acausal $\xi$ carry  the
Hilbert space $\hh(\xi_A)$. When one inquiries about possible
reduced dynamics on it, that means following the evolution for all
possible initial states on $\hh(\xi_A)$, while keeping the reduced
state of the remaining subsystem $\hh(\xi_B)$ and the correlations
between the subsystems fixed.  Then the evolution of a
$d_A$-dimensional state $\rho_A$ will be given by a linear,
possibly non-CP,  map. In particular
\cite{ctz},
under the action of a unitary $U$ on an extended system, the
reduced dynamics of $\rho_A$ is given by an affine map,
 with its linear part being a CP
map and the  traceless part related to the initial correlations
between the system $A$ and the ancilla $B$ \cite{ctz}. More
precisely,
\be
\Xi(\rho_A)=\sum_kM_k\rho_A
M^\dag_k+\sum_{ijl}c_{l,ij}\Gamma_{ij}\sigma_l,
\ee
where the first expression on the rhs is a completely positive map
$\Lambda(\rho_A)$ that is written in Kraus decomposition form, the
coefficients $c_{l,ij}$ depend on $U$ alone, and
\be
\Gamma_{ij}=(\gamma_{ij}-\alpha_i\beta_j)/d_Ad_B,
\label{corrtens}
\ee
 is the correlation
tensor \cite{mah95}, and the matrices $\sigma_l$ are the
generators of SU$(d_B)$.

Any CP map $\Lambda$ can be decomposed into a unital CP map
$\Lambda_0$ and the constant term $\sum_l a_l\sigma_l$. Then if
$\Xi(\1)=\1$, the constant parts of the map cancel, and one  is
left with
\be
\Xi(\rho_A)=\Lambda_0(\rho),
\ee
i.e., with a completely positive evolution. It should be noted
that the requirement of unital evolution of the reduced subsystem
imposes a simultaneous constraint on both allowed global unitaries
$U$ and the initial states.
\section{From global states to local information and back}

Choosing an acausal set is analogous to fixing a leaf in the
foliation in the continuum case. On such a slice it is meaningful
to define reduced density matrices, as in the previous section.
However, an inverse procedure (reconstructing $\rho_\xi$ from
$\rho_{\xi_A}$ and $\rho_{\xi_B})$ is not unique. As it is seen
from Eq.~(\ref{Fano}), an additional information on correlations
is required. For example, in the two-qubit case a generic density
matrix is specified by 15 real parameters, while the reduced
density matrices specify only six of them. Even for a pure state
$\rho_\xi$ there is more information in $\Gamma$ than just the
degree of entanglement. All the maximally entangled states have
$\alpha_i=\beta_j=0$, but are distinguished by different elements
of $\Gamma$.

Dynamics can help in reducing redundancy of the possible state
reconstruction. In addition, change of Lorentz frame that induces
the change of (semi-global) foliations, is analogous to the choice
of a different partition into acausal sets.

If  we consider an evolution of $\xi_A$ that is a subset of the
acausal set $\xi$ into $\zeta_A$ of the acausal set $\zeta$, which
may be also taken as the subsets of the acausal sets $\xi'$ and
$\zeta'$ respectively, the reduced dynamics $\Xi(\rho_A)$ should
be the same. Given the reduced density matrices knowledge of the
local evolution constrains their the possible embedding into the
global states on $\hh(\xi)$ and $\hh(\zeta)$.
 To summarize the situation, the Lorentzian
structure of the space-time is defined through the definition of spacelike foliations of the causal
history and how the quantum states living on these foliations are related by  Lorentz boosts. The
quantum states associated to each spacelike hypersurface are not uniquely determined by the local
information, i.e., the reduced density matrices, but we need the global information contained in
the correlations between these density matrices. These correlations are a priori induced by the
precise dynamics (unitarity operators) on the causal set. At the end of the day, more work is
needed in order to be able to define precisely how the dynamics determines the action of Lorentz
transformations of quantum states on the causal set. This is left for future investigation.

\medskip

 To conclude, we have discussed the quantum causal histories
as introduced in \cite{fotini03, fotini00}. These are basically
causal sets dressed with Hilbert spaces (on the nodes) and
(completely positive) operators on the arrows describing the
evolution of the quantum states along the causal set. To start with,
we have addressed the issue of closed timelike loops and shown that
they carry no relevant extra degrees of freedom. Then we have
noticed that the causal links of generic quantum causal histories
are observer-dependent: they depend on the basis of measurements
chosen by the observer. To get rid of this observer dependence, we
must add causal links reflecting the way the dynamics entangle the
various systems. Moreover, restrictions on the measurement bases
also restrict the allowed unitary evolutions. At this level, it
appears more natural to switch to a dual picture where causal
histories are described in terms of quantum computational networks
(quantum circuits):  at the end of the day, these are the only
causal histories that allow the introduction of  quantum mechanics
without imposing restrictions on the measurements. We have also
discussed the requirement of complete positivity of the evolution
operators. This is quite a restrictive requirement. However, it
 results from a standard posing of the initial-value
problem and the assumption that the maximally mixed state remains
such in the course of evolution. Finally, we discussed the action of
Lorentz boosts on such quantum causal histories and explained that
it involves understanding how the global quantum states can be
deduced from the states (reduced density matrices) of the
subsystems. This is a hard problem which involves understanding how
the causal set dynamics creates correlations between the subsystems
and which we leave as an open issue.

\acknowledgments

We thank Steve Bartlett,  Fotini Markopolou,  Lee Smolin and Bill
Unruh for stimulating discussions. Part of the work by ERL was
performed at the Perimeter Institute.



\begin{thebibliography}{99}

\bibitem{fotini03}
 E.~Hawkins, F.~Markopoulou and H.~Sahlmann,
  Class.\ Quant.\ Grav.\  {\bf 20}, 3839 (2003)
  [arXiv:hep-th/0302111].


\bibitem{fotini00}
  F.~Markopoulou,
  Class.\ Quant.\ Grav.\  {\bf 17}, 2059 (2000)
  [arXiv:hep-th/9904009].



\bibitem{causalSF}
E.~R.~Livine and D.~Oriti,
Nucl.\ Phys.\ B {\bf 663}, 231 (2003) [arXiv:gr-qc/0210064].


\bibitem{nc}  A. Nielsen and I. L. Chuang, {\it Quantum Computation and
Quantum Information\/}  (Cambridge University Press, New York,
2000).

\bibitem{lloyd} S. Lloyd,  Science {\bf 273}, 1073 (1996); S.
Lloyd,   
  arXiv:quant-ph/0501135.


\bibitem{gr} S. W. Hawking and G. F. R. Ellis,  {\it The Large Scale
Structure of Space-Time\/} (Cambridge University, Cambridge,
1973).

\bibitem{rmp}   A.~Peres and D.~R.~Terno,
  Rev.\ Mod.\ Phys.\  {\bf 76}, 93 (2004)
  [arXiv:quant-ph/0212023].


\bibitem{ctz}
H. Carteret, D. R. Terno, and K. \.{Z}yczkowski,
arXiv:quant-ph/0512167.


\bibitem{fn} V. P. Frolov and I. D. Novikov, \textit{Black Hole
Physics: Basic Concepts and New Developments\/} (Kluwer,
Dordrecht, 1998);

M. Viser, \textit{Lorenzian Wormholes\/} (American Institute of
Physics, New York, 1995).

\bibitem{worm} M.  S. Morris and K. S. Thorne, Am J. Phys. {\bf
56}, 395 (1988);

M.  S. Morris, K. S. Thorne and U. Yurtsever, Phys. Rev. Lett. {\bf
61}, 1446 (1988).

\bibitem{deutsch} D. Deutsch, Phys. Rev. D {\bf 44}, 3197 (1991).

\bibitem{bac} D. Bacon, Phys. Rev. A {\bf 70}, 032309
(2004) [arXiv:quant-ph/0309189].

\bibitem{OHrep}
  F.~Girelli and E.~R.~Livine,
  Class.\ Quant.\ Grav.\  {\bf 22}, 3295 (2005)
  [arXiv:gr-qc/0501075].


\bibitem{peres} A. Peres, {\it Quantum Theory: Concepts and Methods\/}
(Kluwer, Dordrecht, 1993).

\bibitem{fotini05}   D.~W.~Kribs and F.~Markopoulou,
  arXiv:gr-qc/0510052.



\bibitem{mah95} J. Schlienz and G. Mahler, \pra {\bf 52}, 4396
(1995).

\bibitem{jeanes} E.T. Jaynes, \textit{Information theory and statistical mechanics},
in \textit{1962 Brandeis Lectures}, vol. 3, ed. by K.W. Ford
(Benjamin, Elmsord, New York 1963) p.~181.


\bibitem{buzek}
M. Ziman, M. Plesch, and  V. Bu\v{z}ek, arXiv:quant-ph/0406088;

M. Ziman, M. Plesch,   V. Bu\v{z}ek, and P. \v{S}telmachovi\v{c},
\pra {\bf 72}, 022106 (2005) [arXiv:quant-ph/0501102].

\end{thebibliography}
\end{document}